\documentclass[twocolumn,letter]{jpsj3}

\def\l{\langle}
\def\r{\rangle}
\def\vecS{\mbox{\boldmath $S$}}
\def\vecD{\mbox{\boldmath $D$}}
\def\vecr{\mbox{\boldmath $r$}}

\def\runauthor{Yusuke Tomita {\it et al.}}

\title{
Monte Carlo Study of Relaxor Systems: \\
A Minimum Model of Pb(In$_{1/2}$Nb$_{1/2}$)O$_3$}

\author{%
Yusuke \textsc{Tomita}\thanks{E-mail address: ytomita@issp.u-tokyo.ac.jp},
Takeo \textsc{Kato},
and Kazuma \textsc{Hirota}$^{1}$
}

\inst{%
Institute for Solid State Physics, University of Tokyo, Kashiwa 277-8581, Japan \\
$^{1}$Department of Earth and Space Science, Faculty of Science, Osaka University, Toyonaka 560-0043, Japan
}

\recdate{\today}

\abst{%
We examine a simple model of Pb(In$_{1/2}$Nb$_{1/2}$)O$_3$ (PIN), which includes
both long-range dipole-dipole interaction and random local anisotropy. An improved algorithm
optimized for long-range interaction has been applied to efficient large-scale Monte Carlo
simulation. We demonstrate that the phase diagram of PIN is qualitatively reproduced
by this minimum model. Some characteristic features of relaxors such as nanoscale 
domain formation, slow dynamics, and dispersive dielectric responses are also examined.}

\kword{%
relaxors, ferroelectric phase transition, antiferroelectric phase transition, 
Monte Carlo algorithm, domain formation, slow dynamics, dispersive
dielectric responses, Pb(In$_{1/2}$Nb$_{1/2}$)O$_3$
}

\begin{document}
\maketitle

Relaxors, whose first discovery dates back to over half a century ago~\cite{Smolenskii58}, have
newly attracted much interest over the past few decades owing to their colossal dielectric and
piezoelectric responses that are appealing for industrial applications. Despite intensive research, however, 
the physical origin of the unusual properties of relaxors is not fully understood yet.~\cite{Bokov06}
This is because nanoscale intrinsic randomness in relaxor systems has to be addressed appropriately.

As a simple example to understand the physics of relaxors, let us choose
the lead-based relaxor Pb(In$_{1/2}$Nb$_{1/2}$)O$_3$ (PIN), which is the target of the present theoretical work.
The intrinsic inhomogeneities of relaxor systems originate from the random configuration of In$^{3+}$ 
and Nb$^{5+}$ at the B-site in the perovskite structure, while the dielectric property of such systems is governed mainly by
the replacement of Pb and O ions. This system has the advantage that the strength 
of randomness can be controlled by adjusting annealing temperature.~\cite{Bokov84} For the lowest annealing temperature,
an alternate order of In and Nb atoms stabilizes the fourfold antiferroelectric (AFE) phase. 
For the higher annealing temperature, an increase in B-site randomness decreases
AFE transition temperature.
Under sufficiently strong randomness in the B-site, the ferroelectric phase 
accompanies relaxor properties. This behavior of PIN suggests that the potential FE instability exists behind the AFE phase,
and that its emergence by the selective suppression of the AFE phase due to B-site randomness is relevant
to FE relaxors. Recent experimental progress in the X-ray diffraction analysis
facilitated by strong-intensity photon sources has also enabled us to detect the potential competition between the FE and
AFE phases of PIN from the viewpoint of lattice dynamics.~\cite{Ohwada08,Ohwada10} 

There are several theoretical methods for elucidating the emergence of relaxor properties.
They utilize the existing effective theory dealing with a strong randomness
such as the extension of the Ginzburg-Landau-Devonshire theory,~\cite{Devonshire54,Liu06}
the spherical random-bond-random-field model,~\cite{Blinc99}
and the dynamical-related model.~\cite{Bell93,Vugmeister98}
Although these approaches have provided useful description of relaxor properties, 
it is unclear how their model parameters should be derived microscopically.
At present, first-principles calculation is playing
a crucial role in the microscopic description of ferroelectrics and ferroelectric relaxors.
In the pioneering work by Cohen and coworker~\cite{Cohen92a,Cohen92b},
the electron-structure calculation of BaTiO$_3$ and PbTiO$_3$ indicated the importance
of the covalent nature among composite atoms through Ti-O hybridization.
After the success of their work, microscopic first-principles approaches have been adopted in a number of issues.
Further theoretical progress has been achieved by a hybrid method combining the microscopic derivation of
the effective Hamiltonian and its numerical simulation. Zhong \textit{et al.} 
have derived the effective Hamiltonian of BaTiO$_3$
by extracting the adiabatic potential of important phonon modes 
from first-principles calculation.~\cite{Zhong94, Zhong95}
Monte Carlo simulation for this effective Hamiltonian has enabled the
successful reproduction of the sequential finite-temperature phase transitions
of BaTiO$_3$ in good agreement with experimental results. Recently, the hybrid method has also been applied
to the study of relaxor systems.~\cite{Tinte06,Burton08}

Thus, the utilization of first-principles calculation is a promising method for understanding
relaxors. However, there still remain several difficulties to overcome for the actual application of this method
to the study of relaxors. One major difficulty is in the use of the numerical solver of the effective Hamiltonian 
for phonons. Even though the effective Hamiltonian derived for relaxors is in a simple form, 
its numerical simulation may become extremely difficult because it is inevitable to encounter
slow dynamics inherent to random systems. The Monte Carlo
simulation of the effective Hamiltonian needs a large number of Monte Carlo steps owing to its long correlation time.
Therefore, the development of appropriate numerical solvers for relaxor systems seems indispensable.
This situation may remind us of Monte Carlo studies of spin-glass systems. Although the Hamiltonian
of spin-glass systems has a simple form, much effort has to be exerted for sufficient Monte Carlo sampling
to understand their peculiar slow dynamics. In the research field of relaxors, however, the importance of
developing a numerical solver has not been addressed so far.

In this paper, we propose a simple model of ferroelectric relaxors
to explain the phase diagram of PIN. Our model includes both 
long-range dipole-dipole interaction and local randomness. 
We apply the new effective algorithm optimized for long-range interaction
proposed by Fukui and Todo.~\cite{Fukui09} This new algorithm enables us 
to simulate long-range interaction systems with the cost of O($N$) with respect to
the system size $N$, while the conventional simulation for the same system
takes the cost of O($N^2$). In addition to the substantial reduction in
computational time by employing this algorithm, we adopt the exchange
Monte Carlo method to attenuate the slow dynamics of random systems.~\cite{Hukushima96}
We show that our simple model may qualitatively reproduce 
the phase diagram of PIN. Moreover, we demonstrate that some characteristic features of relaxors
such as domain formation and dispersive dielectric response are reproduced reasonably.
Our results indicate the power of the sophisticated Monte Carlo method as well as the possibility
that relaxor systems may be represented by a simple model.

We consider the model Hamiltonian for PIN on a 2D square lattice as
\begin{eqnarray}
{\cal H} &=& \sum_{i < j}\left[\frac{\vecS_i \cdot \vecS_j}{r^3_{ij}}
-3\frac{(\vecr_{ij} \cdot \vecS_i)(\vecr_{ij} \cdot \vecS_j)}{r^5_{ij}}\right] \nonumber \\
&-& \sum_i (\vecD_i \cdot \vecS_i)^2,
\end{eqnarray}
where $\vecS_i$ is a unit vector in the $xy$-plane representing the dipole moment
on the $i$th unit cell induced by the off-center replacement of the Pb atom.
The first term of the Hamiltonian
is the dipole-dipole interaction dependent on the relative position
$\vecr_{ij} = \vecr_j - \vecr_i$ between the sites $i$ and $j$, while
the second term describes local anisotropy whose direction and strength are
denoted as $\vecD_i$. In order to reproduce the phase diagram of PIN, 
we design our model such that the FE phase is stabilized by dipole-dipole interaction,
while the AFE phase is stabilized in the presence of an alternative change in $\vecD_i$.
Supposing that B-site randomness affects only the local energy change through
the anisotropy parameter $\vecD_i$, we can expect that all the features
of PIN, i.e., the AFE transition in the ordered PIN, its suppression
by B-site randomness, and the appearance of the FE domain in the sufficiently
disordered PIN are reproduced.

The detailed setting of our model is given as follows.
In a naive 2D square lattice, the dipole-dipole interaction does not lead to
ferroelectric instability because the columnar antiferroelectric state is more favored 
in orthogonal lattices.~\cite{Tomita09}
Therefore, we modify our model slightly: We divide the
unit cells into two bipartite sublattices named P and Q, and shift
only the Q-sites in the $z$-direction by a unit length.
This rearrangement of unit cells ensures ferroelectric instability
in the absence of local anisotropy. The antiferroelectric instability 
is driven by an alternative arrangement of two types of anisotropy, 
namely, $\vecD_1$ and $\vecD_2$ in the P- and Q-sites, where 
$\vecD_1 = (D/\sqrt{2}, -D/\sqrt{2})$ and $\vecD_2 = (D/\sqrt{2}, D/\sqrt{2})$.
The randomness of $\vecD$ is controlled by the probability $p$, where
$\vecD_1$ ($\vecD_2$) is attached to the P-(Q-)site with a
probability $(1-p)^2$, oppositely to $p^2$,
and is turned off ($\vecD$={\bf 0}) with $2p(1-p)$. The two limiting values $p=0$ and
$p=1/2$ correspond to the ordered and completely disordered PINs, respectively.

The model Hamiltonian is examined by Monte Carlo simulations
under a periodic boundary condition to minimize the strong surface effect due to long-range interaction.
Long-range interactions are summed up by the Ewald summation technique.~\cite{Weis03}
We employ the ${\rm O}(N)$ method based on Walker's algorithm 
for efficient update~\cite{Fukui09,Walker77} as well as the temperature-exchange algorithm.~\cite{Hukushima96}
One exchange trial between replicas was made for each of the $10$ MC steps.
The system size is taken up to $N = 32 \times 32$, for which 
$3\times 10^6$ MC steps for thermalization and $2\times 10^5$ MC steps for measurement 
were needed in the severest case of $p=1/2$.
The sample average is taken over $10$ different random configurations of $\vecD_i$ 
for each $p$. Throughout this paper, the strength of the anisotropy is fixed
as $|\vecD_i| = 1$.

\begin{figure}[tb]
\begin{center}
\includegraphics[width=7cm]{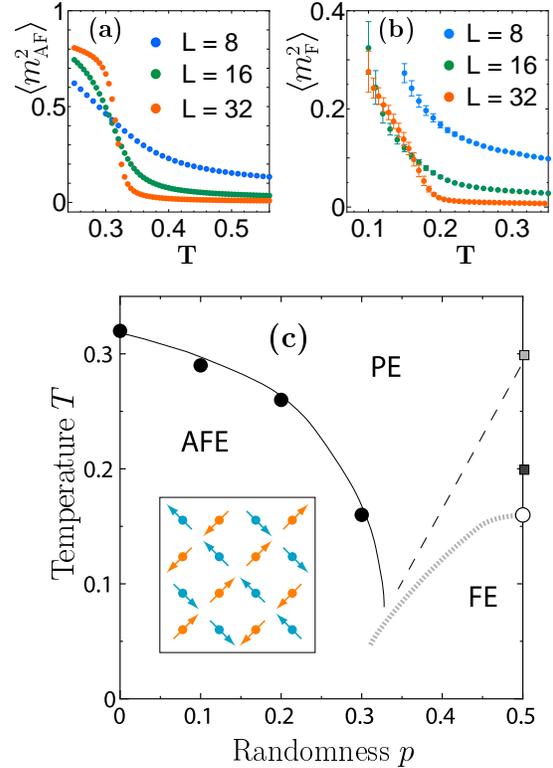}
\end{center}
\caption{
(Color online) (a) A plot of four-fold antiferroelectric(AFE) order parameters in the absence of randomness ($p=0$) 
and (b) a plot of ferroelectric(FE) order parameters at $p=0.5$ are shown for 
different system sizes $N=L\times L$. 
(c) Phase diagram obtained by the minimum model.
Phase boundaries in the figure are visual guides. }
\label{fig:phase_diagram}
\end{figure}

Figure~\ref{fig:phase_diagram}(a) shows the temperature dependence
of the average squared four-fold staggered polarization $m_{\rm AF}$ for the ordered PIN ($p=0$).
The pattern of the AFE ordering, which is shown in Fig.~\ref{fig:phase_diagram}(c), agrees with
experimental observations.
As the system size $N = L \times L$ increases, a sharp increase in the order parameter
appears below a critical temperature, indicating the phase transition.
The transition point is determined by the crossing point of the Binder parameters,
$\l m_{\rm AF}^4 \r/\l m_{\rm AF}^2 \r^2$, of $L=16$ and $32$.
By a similar analysis of nonzero values of $p$, the phase boundary of AFE in
the $p$-$T$ plane is determined, as shown by full circles in Fig.~\ref{fig:phase_diagram}(c).
For small values of $p$, the transition temperature of AFE is suppressed by
the B-site randomness. 
For sufficiently large values of $p$, the FE domain develops at low temperatures
instead of the AFE phase. Figure~\ref{fig:phase_diagram}(b) shows
the average squared uniform polarization as a function of temperature 
in the completely disordered case ($p=1/2$). An abrupt increase in squared polarization
below a threshold temperature for the $N=32 \times 32$ system indicates 
a rapid development of the FE domain.
In our simulation, however, no long-range FE ordering could be detected by finite-size
scaling because of the very slow relaxation of the Monte Carlo sampling and
the complex size dependence in the presence of strong randomness.
Here, we roughly estimate the threshold temperature at which the local FE instability rises up
by Binder-parameter analysis between $L=16$ and $32$, and plot it 
using the empty circle in Fig.~\ref{fig:phase_diagram}(c). 
At approximately $p = 0.4$, the competition between AFE and FE makes the Monte Carlo calculation severe,
and the phase boundary could not be determined. Rough interpolations for AFE and FE are 
shown by the solid curves in Fig.~\ref{fig:phase_diagram}(c) only as visual guides.
The obtained phase diagram is in good agreement with experimental results of PIN.

\begin{figure}[tb]
\begin{center}
\includegraphics[width=7cm]{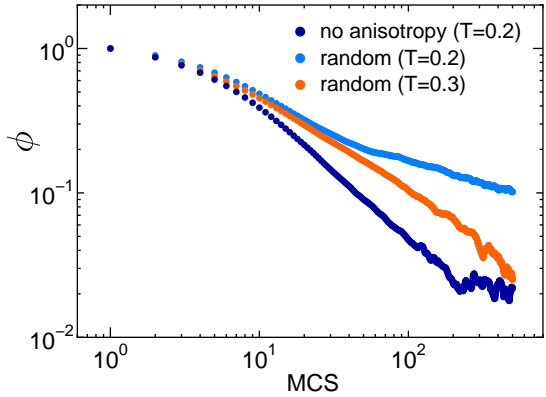}
\end{center}
\caption{
(Color online) Log-log plot of the normalized time autocorrelation functions.
While the autocorrelation functions decay smoothly
for the no-anisotropy system at $T=0.2$ and the random system at $T=0.3$,
the autocorrelation function shows a slow decay for the random system
at $T=0.2$.}
\label{fig:auto_correlation}
\end{figure}

For large values of $p$, some characteristic features of relaxor systems appear
below a crossover temperature, which locates higher than the threshold temperature
for FE domain formation,
as indicated roughly by the dashed line in Fig.~\ref{fig:phase_diagram}(c).
The emergence of relaxor properties is indicated significantly by the slow convergence of
the Monte Carlo average. Figure~\ref{fig:auto_correlation} shows the normalized
autocorrelation of the energy $E$ for $N = 32 \times 32$ defined by
\begin{equation}
\phi(t) = \frac{\l E(0)E(t)\r - \l E\r^2}{\l E^2\r - \l E\r^2}
\end{equation}
as a function of the Monte Carlo step $t$ in the completely random case ($p=1/2$).
Here, for the purpose of examining the dynamics of the system, we turned off
the exchange process between replicas.
We find that the decay of the autocorrelation function is extremely slow at $T=0.2$
in clear contrast with that at a higher temperature $T=0.3$.
This slow energy relaxation indicates a glasslike behavior for relaxors at low temperatures.
In order to confirm that this slow relaxation comes from randomness of the B-site,
we also calculated the autocorrelation at $T=0.2$ in the absence of local anisotropy ($|\vecD_i| = 0$),
for which normal FE ordering is expected. As seen in Fig.~\ref{fig:auto_correlation},
the autocorrelation function follows a simple decay with no anomalous slow relaxation.
Thus, the present model is expected to exhibit relaxor properties under sufficiently 
strong randomness below a crossover temperature.

\begin{figure}[tb]
\begin{center}
\includegraphics[width=7cm]{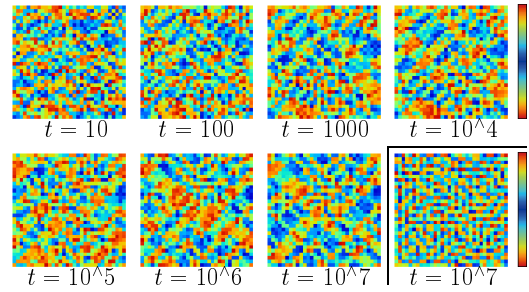}
\end{center}
\caption{
(Color online) Snapshots of dipole configuration for $p=0.5$ and $T=0.1$.
Each color denotes the angle of dipole moment. Reference color bars are
shown on the right. The bottom color denotes $-\pi$, and the top color
denotes $\pi$. The last graph shows the result for the same parameters
but for a model with a dipole interaction interrupted up to the next-nearest sites.}
\label{fig:snap_shot}
\end{figure}

In order to visualize the glassy state realized in the relaxor phase,
we show in Fig.~\ref{fig:snap_shot} the spatial pattern of dipole directions by taking snapshots
as a function of the MC steps for $p=0.5$ and $N = 32\times 32$.
We can see that mosaic-like FE domains are formed after a finite relaxation time.
The boundaries (domain-wall regions) between the neighboring FE domains are rather clear.
One may consider that the FE domain structure is determined by a partially ordered
configuration of anisotropy, the so-called chemical nanoregion (CNR). The present FE domain is, however,
irrelevant to CNR, because partial ordering is absent for a complete random
choice of anisotropy at each site in our model.

As has already been mentioned, the present system possesses
FE instability in the absence of local anisotropy (${\boldmath D}={\boldmath 0}$).
The random configuration of anisotropy similarly causes a local FE instability due to 
effective cancellation of anisotropy, but at the same time prevents the growth of an FE domain 
into a uniform FE order, as seen in Fig.~\ref{fig:snap_shot}.
A small but finite change of the snapshot in the long-time region of the Monte Carlo simulation
indicates that a very slow fluctuation of the FE domain governs the relaxor properties in 
the present model. Here, we should note that the evolution of the dipole configuration along
the Monte Carlo step is not directly related to actual real-time dynamics. The snapshots, however,
provide intuitive and fruitful insights for understanding physical processes in relaxor systems.
In our model, the long-range nature of the dipole interaction is essential
for the formation of the FE domain structure. To see this, we show a snapshot of 
Monte Carlo simulation for a model with dipole interaction
interrupted up to the next-nearest site in the last graph of Fig.~\ref{fig:snap_shot}.
We find that the domain structure disappears in this modified model.

\begin{figure}[tbh]
\begin{center}
\includegraphics[width=7cm]{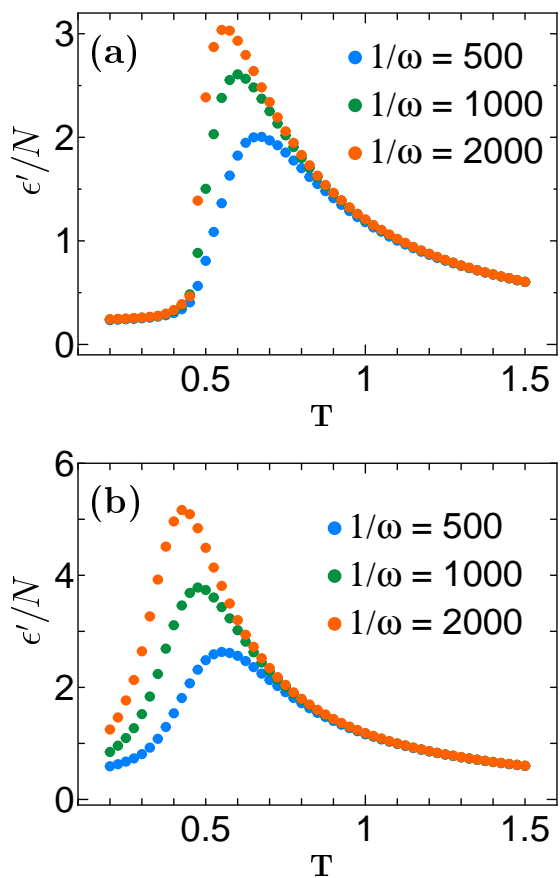}
\end{center}
\caption{(Color online) Dielectric constants evaluated as functions of temperature under
the ac electric field for several frequencies in the (a) ordered case ($p=0$) and 
(b) completely disordered case ($p=1/2$).}
\label{fig:dielectric}
\end{figure}

The existence of nanoscale frustrated FE domains indicates
the high degeneracy of glassy states. This degeneracy is expected to be responsible for 
large dielectric responses since a significant change in the polarization of such FE domains
is possible by a small external electric field.
To see this, we measure the Monte Carlo evolution of the total polarization 
under a small `ac' electric field, which varies periodically along the Monte Carlo step.
We then calculate the `ac' dielectric constant in the linear response regime.
We should note again that the `ac' dielectric constant thus obtained is not equal to
the real-time ac dielectric constant. Furthermore, we should keep in mind that this is
a nonequilibrium response since the relaxation to equilibrium states is never realized
in the glassy phase in the present simulation. Nevertheless, the `ac' dielectric response 
calculated here is expected to mimic the actual ac response in a qualitative level. In Fig.~\ref{fig:dielectric},
we show the `ac' dielectric constant as a function of temperature
for three frequencies in the ordered ($p=0.0$) and completely disordered ($p=1/2$) cases.
In both cases, dielectric constant shows a peak at approximately the transition (crossover) 
temperature. The marked difference is found in the low-temperature phase.
In the ordered case, dielectric constant sharply drops in the low-temperature
AFE phase, and becomes almost independent of frequency. 
On the other hand, in the disordered case, it decreases gradually 
as temperature decreases, and a strong frequency dependence remains.
This strong dispersion of dielectric constant at low temperatures 
can be explained as follows. Under a high-frequency electric field, 
each dipole inside FE domains may respond to an external field.
Under a low-frequency electric field, however, those with frustration that construct large domains
start to respond. Therefore, in a minimum model, a dipolar glass with ferroelectric
ordering is realized. This ordering causes a broadened dielectric constant
and a strong dependence on frequency in the relaxor phase.

In summary, we examined a simple theoretical model of PIN 
composed of dipolar interaction and local random anisotropy.
We demonstrated that efficient Monte Carlo simulations equipped with an improved
algorithm optimized for long-range interaction may access several
characteristic features of relaxors. The phase diagram of PIN was qualitatively reproduced
by appropriate inclusion of the intrinsic competition between the AFE and FE phases.
By the examination of the Monte Carlo evolution of the dipole configuration, 
we demonstrated some of the glassy behaviors inherent to relaxor systems such as
the FE domain formation, extremely slow dynamics, and strong frequency dependence
of dielectric responses.

We end this paper by mentioning the future outlook of theoretical approaches to studying relaxors.
The model we treated here includes a few artificial assumptions. They, however, will be removed
by replacing our model with the effective Hamiltonian derived by first-principles calculation.
We stress that the smart Monte Carlo algorithm is applicable not only to the rotator model
but also to the continuous-variable model. The hybridization of the first-principles calculation and
statistical approach based on the Monte Carlo simulation will be an effective means of
elucidating the microscopic origin of relaxors.

The computation in the present work is executed on computers at the Supercomputer
Center, Institute for Solid State Physics, University of Tokyo. The present work 
is financially supported by a MEXT Grant-in-Aid for Scientific Research (B) 
(19340109), by a MEXT Grant-in-Aid for Scientific Research on Priority Areas ``Novel 
States of Matter Induced by Frustration'' (19052002), and by the Next Generation 
Supercomputing Project, Nanoscience Program, MEXT, Japan

\end{document}